\documentclass[sigchi]{acmart}
\graphicspath{{figures/}{pictures/}{images/}{./}} 
\usepackage{textcomp}                  
\usepackage[utf8]{inputenc}
\usepackage{amsmath}
\usepackage{mathtools}
\usepackage{amssymb}
\usepackage{mathrsfs}
\usepackage{balance}
\usepackage{tabu}                      
\usepackage{booktabs}                  
\usepackage[ruled,vlined,linesnumbered]{algorithm2e}
\usepackage{hyperref}
\usepackage{gensymb}
\usepackage{physics}

\newif\ifdebug
\debugtrue

\newif\ifdoubleblind

\newcommand*{\origrightarrow}{}

\renewcommand*{\textrightarrow}{\fontfamily{cmr}\selectfont\origrightarrow}

\newif\ifdebug
\debugtrue
\newif\ifdoubleblind
\doubleblindtrue

\makeatletter
\newcommand*{\defeq}{\mathrel{\rlap{%
                     \raisebox{0.3ex}{$\m@th\cdot$}}%
                     \raisebox{-0.3ex}{$\m@th\cdot$}}%
                     =}
\makeatother

\copyrightyear{2019}
\acmYear{2019}
\setcopyright{acmcopyright}
\acmConference[CHI 2019]{CHI Conference on Human Factors in Computing Systems
Proceedings}{May 4--9, 2019}{Glasgow, Scotland Uk}
\acmBooktitle{CHI Conference on Human Factors in Computing Systems Proceedings
(CHI 2019), May 4--9, 2019, Glasgow, Scotland Uk}
\acmPrice{15.00}
\acmDOI{10.1145/3290605.3300643}
\acmISBN{978-1-4503-5970-2/19/05}

\DeclareMathAlphabet      {\mathbfit}{OML}{cmm}{b}{it}

\newcommand{\MI}{\left[ i \right]}
\newcommand{\MJ}{\left[ j \right]}
\newcommand{\MB}{\left[ b \right]}
\newcommand{\MII}{\left[ i-1 \right]}
\newcommand{\LEFT}{\textrm{left}}

\newcommand{\TOP}{\textrm{top}}

\newcommand{\WIDTH}{\textrm{width}}
\newcommand{\HEIGHT}{\textrm{height}}

\begin{document}
\title{ORC Layout: Adaptive GUI Layout with OR-Constraints}

\author{Yue Jiang}
\affiliation{
  \institution{Department of Computer Science\\University of Maryland, College Park, MD, USA}
}
\email{yuejiang@cs.umd.edu}

\author{Ruofei Du}
\affiliation{
  \institution{Google LLC, San Francisco, CA \&\\
  University of Maryland, College Park, MD, USA}
}
\email{me@duruofei.com}

\author{Christof Lutteroth}
\affiliation{
  \institution{Department of Computer Science\\University of Bath, Bath, United Kingdom}
}
\email{c.lutteroth@bath.ac.uk}

\author{Wolfgang Stuerzlinger}
\affiliation{
  \institution{School of Interactive Arts + Technology (SIAT)\\Simon Fraser University, Vancouver, BC, Canada}
}
\email{w.s@sfu.ca}

\renewcommand{\shortauthors}{Jiang Y., Du R., Lutteroth C., and Stuerzlinger W.}

\begin{abstract}
We propose a novel approach for constraint-based graphical user interface (GUI) layout based on OR-constraints (ORC) in standard soft/hard linear constraint systems. ORC layout unifies grid layout and flow layout, supporting both their features as well as cases where grid and flow layouts individually fail. We describe ORC design patterns that enable designers to safely create flexible layouts that work across different screen sizes and orientations. We also present the ORC Editor, a GUI editor that enables designers to apply ORC in a safe and effective manner, mixing grid, flow and new ORC layout features as appropriate. We demonstrate that our prototype can adapt layouts to screens with different aspect ratios with only a single layout specification, easing the burden of GUI maintenance. Finally, we show that ORC specifications can be modified interactively and solved efficiently at runtime.
\end{abstract}

%
%



\begin{teaserfigure}
  \includegraphics[width=\textwidth]{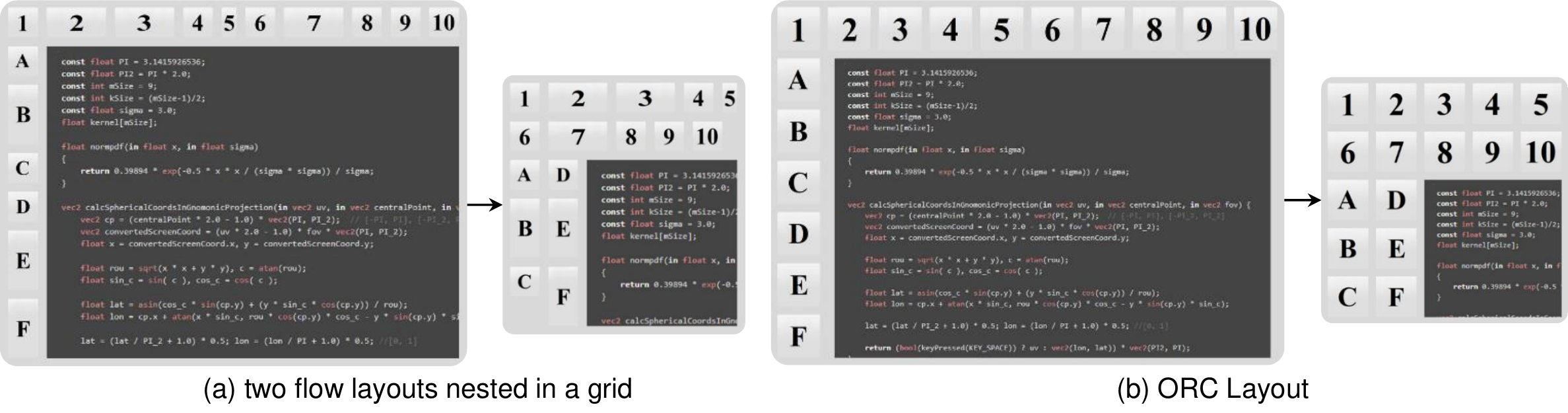}
  \caption{Examples of GUIs combining elements of grid and flow layouts, which can only be realized with ORC layouts. Both (a) and (b) show the results before and after resizing the window. (a) shows a traditional solution with two flow layouts nested in a grid. It looks irregular since the original sizes of the widgets in the toolbars are not the same. Using ORC layouts in (b), we can add constraints that cross-cut the GUI hierarchy, ensuring that the tool buttons all have the same size and hence consistent appearance.}  
  \Description{The teaser image shows an example of a layout that can only be realized with an ORC layout. }
  \label{fig:teaser}
\end{teaserfigure}

\maketitle

\section{Introduction}

Automatically computed layouts are widely used in many applications, including webpages, graphical user interfaces (GUIs), documents, slides, and mobile apps. WYSIWYG (``what you see is what you get'') GUI editors enable designers to create layouts by defining the positions and sizes of widgets, as well as specifying how a layout adapts to changes, through layout models.

Grid layout models such as grid-bag can adjust the relative size of widgets, while keeping mutual alignments intact. Flow layout models can break widgets into new rows or columns, but are limited in the way widgets can be aligned. Constraint-based layout models, the most powerful option, are becoming widespread and can be used to specify any grid layout, can align widgets across different groups, and relax constraints between aligned widgets to enable more flexibility. For example, Apple’s AutoLayout \cite{sadun2013ios} supports adaptive interfaces on devices ranging from desktops to smartphones. CSS's Flex(ible) Box\footnote{CSS Flexible Box: \url{https://w3schools.com/css/css3_flexbox.asp}}, widely supported in web browsers, applies constraints for dynamically fitting the content and solving alignment problems. 

Despite their power, current constraint-based layout models have limitations. First, they do not support flow layouts. Second, for GUIs to work well in both portrait and landscape orientations and on small as well as large screens, separate layout specifications have to be created for each case and activated on demand. Current constraint-based layout models are not flexible enough to deal with all such changes directly, which increases the workload of the designer: several layout specifications have to be created and maintained. During the evolution of a GUI, designers then have to manually synchronize changes between different layout specifications. If all cases could be covered in a single layout specification, this could make GUI design more efficient and less error-prone. 

Specifying flexible layouts with constraints is not easy, and specifications are typically formulated as a linear programming problem, through a system of linear equations and inequalities, which are either hard or soft constraints. A dedicated solver then calculates the position and size of each widget. A specification can be ambiguous and lead to unpredictable results, or be conflicting and have no result at all. Such problems may be ``hidden'' in a specification and only show in some cases, {\em e.g.}, for a particularly small screen. 

In this paper, we propose ORC layouts -- a novel approach to constraint-based layout that supports OR-constraints in a standard soft/hard linear constraint system.
An OR-constraint is a disjunction of multiple constraints (only one part needs to be true to satisfy the OR-constraint) where the whole OR-constraint is a hard constraint and each part is a soft one. Our method allows designers to specify a layout that can adapt to drastically different screen sizes and orientations using a single layout specification. A GUI can then be adapted to changes at runtime without having to consider several separate specifications, avoiding inefficiencies and inconsistencies. ORC layout also unifies the -- so far separate -- grid and flow layout models, addressing both of their shortcomings. Additionally, it offers some entirely new features that are currently unsupported by other layout models, such as optional widgets.

ORC layouts can be constructed interactively with our ORC Editor. It enables designers to instantiate templates for the ORC layout patterns presented here, and adjust them by modifying their parameters. The ORC editor then automatically maintains all low-level constraints, so designers do not need to deal with them directly.

\autoref{fig:teaser} (b) shows an example that combines elements of grid and flow layouts, only achievable through ORC layouts. On a large screen, the GUI has a toolbar at the top and one on the left, with an edit pane taking up most of the space. For a small screen, the toolbar at the top can break into two rows and the left one into two columns, using a horizontal and a vertical flow layout. While similar behaviour can be achieved with two flow layouts nested in a grid, the results would look different if the original sizes of the toolbar widgets are not the same (\autoref{fig:teaser} (a)). With OR-constraints we can cross-cut the GUI hierarchy, ensuring that the tool buttons all have the same size and hence consistent appearance. Many other layout functionalities can be realized with OR-constraints, including balanced flow layouts and layout alternatives.

\subsection{Contributions}
We contribute two new ideas for constraint-based GUI layout:

\begin{enumerate}
    \item We present ORC layouts, a new powerful way to express GUI layouts, which seamlessly integrate OR-constraints into standard constraint systems for layout specifications. Various layout patterns are supported. One of the biggest benefits is that OR-constraints can unify flow layouts and grid layouts, which opens up options for layouts that work across screen rotations and cases where grid and flow layouts individually fail. 
    \item We present a new approach to solving constraint systems with soft/hard linear and OR-constraints at interactive rates. When resizing a canvas at runtime, our algorithm solves the constraint system based on the previous results. Thus, we can solve a system with 175 constraints in less than a second for operations such as insert, delete, move, and resize.
\end{enumerate}

\section{Related Work}
Our research draws from several related ideas in GUI layout, such as systems of soft/hard linear constraints, minimum/preferred/maximum constraints, CSS, layout solvers, and GUI builders, which we outline in the following.

\subsection{Layout Models}

Layout models are widely used to specify the layout of user interfaces in a way that is (to a certain degree) size independent. Layout managers then determine at runtime the actual GUI layout based on the constraints and the context, such as window size. See also an overview of early approaches \cite{myers1995user,myers2000past}.
Simple layout models, such as group, table, grid, and grid-bag layouts, align widget groups horizontally or vertically and often nest layout containers hierarchically. 
While intuitive, these models often implicitly add more constraints than desired, which introduces maintenance issues \cite{lutteroth2006user,zeidler2012comparing}. They cannot align widgets across a hierarchy, a problem that constraints can address \cite{lutteroth2008modular}. Similar to text layout, flow layouts break widgets into new rows or columns.
Object-oriented models such as Amulet \cite{myers1997amulet} allow developers to combine some properties of flow and grid layouts programmatically.

More modern layout models use constraints, through varying constraint types and explicit or implicit constraint specification.
Linear constraint systems \cite{badros2001cassowary,bill1992bricklayer,borning1997solving,hosobe2000scalable,lutteroth2008domain} are powerful and used today fairly widely, {\em e.g.}, in Apple’s Auto Layout \cite{sadun2013ios}. They use sets of equalities and inequalities to specify absolute and relative alignment of widgets
as well as absolute and relative widget sizes.
Unlike simpler layout models, 
linear constraint systems have a well-understood mathematical basis, which makes results less dependent on implementation details. Linear constraint systems can be combined, allowing reuse of existing specifications in a modular manner \cite{lutteroth2008modular}. Most layout models can be reduced to linear constraint systems \cite{weber2010reduction,zeidler2017tiling}, with the exception of flow layouts. Our work addresses this shortcoming.

Except for flow layouts, existing layout models support only static widget topologies: widget cannot change their position relative to widgets they are aligned with. Flow layouts enable such repositioning in a limited manner: widgets can overflow into a new row or column. Our ORC layouts permit designers to specify flexible layouts that mix grid and flow layouts and seamlessly integrate multiple layout alternatives.

\subsection{Soft/Hard Linear Constraint Systems}

Soft/hard linear constraint systems 
can differentiate between crucial layout characteristics ({\em e.g.}, that a widget must have a certain size to be usable) and merely desirable ones ({\em e.g.}, that two buttons should have the same size).
Two types of constraints express this: hard constraints, which must be satisfied, and soft ones, which can be neglected when it is impossible to determine a solution that can satisfy all constraints in the system.
Further, not all soft constraints should be treated equally, as some soft constraints are more important than others \cite{borning1992constraint}. Hence, a weight is used to rank each soft constraint, with higher weights being equivalent to higher priorities. Hard constraints can be considered as soft ones with infinite weight. 


The intrinsic sizes of each widget are typically defined by min/max/preferred size constraints. The minimum usually depends on the contents of the widget and the maximum is normally limited by the window or screen size, while the preferred size is pre-defined, often by the designer of the GUI editor, or customized by the UI designer in the editor.

\subsection{Layout Solvers}

Various solvers exist for linear constraint-based GUI specifications \cite{badros2001cassowary,bill1992bricklayer,borning1997solving,hosobe2000scalable, zeidler2013evaluating}. Many of them use linear or quadratic programming \cite{badros2001cassowary,bill1992bricklayer,borning1997solving,lutteroth2008domain}, {\em i.e.}, use a linear or quadratic objective function to take constraint priorities into account. Such objective functions penalize deviations from desired positions and sizes, which can improve aesthetics upon resizing \cite{zeidler2012constraint}. Yet, these solvers will try to satisfy \emph{all} constraints simultaneously, subject to priorities in case of conflicts. But they cannot express alternatives, as they support only conjunctions of constraints (AND), but do not permit disjunctions (OR).

Z3 \cite{de2008z3} is a very powerful satisfiability modulo theory (SMT) solver. It supports linear constraints with priorities, other types of constraints, and first-order logic, and is thus able to solve specifications with disjunctive constraints. Using incremental solving, it can perform better if the current constraint system is similar to the previous one. Z3 has previously been suggested for solving formalized CSS layout specifications \cite{panchekha2016automated} and to support layout editing with direct manipulation of individual constraints \cite{hottelier2014programming}. Yet, Z3 has not been used for general GUI/widget layout with OR-constraints before.

\subsection{GUI Builders/Editors}

GUI builder tools support the interactive specification of widgets and layouts and directly generate the corresponding layout models and constraints.
GUI builders need to enable designers to directly and quickly specify complex, flexible layouts, while preventing layout errors. Their set of editing operations determine the range of layouts and resize behaviours that can be specified by a designer.
Core challenges include: 1) powerful layout models support constraints which can be difficult to visualize intuitively, and 2) layout specifications may contain errors that manifest only for particular GUI sizes.

FormsVBT \cite{avrahami1989two} used both textual and visual layout representations, with parallel editing. Gilt \cite{hashimoto1992graphical} introduced reusable styles and visual tabs to simplify layout and appearance specification. Opus \cite{hudson1990interactive} supported visual manipulation of simple layout constraints. Gleicher \cite{gleicher1993graphics} proposed a differential editing approach, where constrained graphical objects can be manipulated directly and changes are continuously reflected in all objects. Unidraw \cite{vlissides1991unidraw} showed how direct manipulation can be implemented by simulating widgets. 

Specifying individual layout constraints can be cumbersome. Thus, some have proposed to infer simple constraints ``by demonstration''. Rockit \cite{karsenty1993inferring} infers constraints on 2D graphical objects based on their locations. Peridot \cite{myers1986creating}, Druid \cite{singh1990druid}, and Lapidary \cite{zanden1991lapidary} generate code based on user interface drawings and example interactions.

Recent GUI builders have focused on creating appropriate layout constraints quickly and robustly through direct manipulation. The Intui builder \cite{scoditti2009new} allows designers to quickly specify the resize behaviour of a GUI by aligning parts of the GUI and designating them either as rigid ``struts'' or flexible ``springs''. The Auckland Layout Editor \cite{zeidler2013auckland} enables designers to specify a constraint-based layout, while automatically keeping the specification solvable and free of overlap. This work also used multiple previews to directly visualize how a layout looks at distinct window sizes. GUI programming by manipulation \cite{hottelier2014programming} also permits editing of individual layout constraints. Designers can interactively disable and enable individual constraints to resolve resulting ambiguities. 

None of the above GUI builders allow designers to specify layouts with OR-constraints. 

\subsection{Layout Alternatives and Alternative Generators}

Linear constraint layouts provide some flexibility for resizing a GUI. However, if a GUI needs to be rendered on a large variety of devices, then a common solution is for the designers to define several layout alternatives, and to select dynamically among them, depending on the rendering context, ({\em e.g.}, available screen size, resolution, and aspect ratio). This approach is common for mobile apps \cite{sahami2013insights,zeidler2017automatic} and ``responsive web design''
\cite{marcotte2011responsive}. For mobile apps, the class of device is used to choose among separate layout specifications, while for responsive web design CSS media queries 
can be specified to apply different layout alternatives, {\em e.g.}, based on screen size.
While flexible, the CSS framework does not offer the same capabilities as ORC layouts, without constructing layouts programmatically with JavaScript.
For example, CSS layouts often use a ``fluid grid'', {\em i.e.}, a rigid grid-like layout in which elements can flow over into separate rows.
Yet, one cannot direct particular widgets to flow in a specific way, as in the connected layout pattern (\autoref{fig:connected_cross_cutting}).
Unlike the ORC approach, a GUI designer still has to manually specify several exact layout alternatives as well as the rules that determine when each alternative should be used.

To automatically improve the user experience based on device properties, tasks and user requirements, some work proposed GUI adaptation, {\em i.e.}, the automatic modification of GUIs based on specific criteria \cite{weld2003automatically}. Fogarty et al. \cite{fogarty2003gadget} used cost functions to generate improved GUIs. SUPPLE \cite{gajos2004supple,gajos2010automatically} uses optimization to generate adapted user interfaces, {\em e.g.}, by replacing widgets or grouping them differently in containers, and also to personalize interfaces for users with disabilities \cite{gajos2008improving}.
Automatic layout generation tools generate proposed alternatives for a given layout \cite{sinha2015responsive,zeidler2017automatic}. Then, designers only need to choose and modify the generated suggestions. A complementary approach is adaptive layout templates \cite{Jacobs2003document}, which also can generate layout alternatives.
Overall, designers are often limited to the (few) alternatives that such tools provide.

In contrast to previous work, our work aims to enable designers to directly specify flexible GUI layouts.
With ORC layouts, designers can leave it up to the solver to determine when to apply a (or a combination of) layout alternative(s), based on screen space or other desirable criteria. Designers can also prioritize, combine, and conditionally use alternatives. For example, a good ORC layout for a smallish medium-sized screen might combine alternatives for medium and small screens, {\em e.g.}, compress only the toolbar but not other UI parts.
ORC layouts also can synchronize layout changes with constraints, {\em e.g.}, to ensure that several rows/columns are broken simultaneously and in a similar manner to achieve a balanced appearance (\autoref{fig:balanced}), even if they are not next to each other.

\setlength{\abovedisplayskip}{-10pt}%
\setlength{\belowdisplayskip}{0pt}%
\setlength{\abovedisplayshortskip}{-10pt}%
\setlength{\belowdisplayshortskip}{0pt}%

\section{OR-constrained (ORC) Layouts}

Here we describe how OR-constrained (ORC) layouts are defined, edited, and solved.

\subsection{ORC Layout Specifications}
In ORC layouts, and as in other constraint-based layouts, the min/max/preferred size of each widget and the relationships among widgets are all defined by constraints. In addition to hard and soft constraints, the constraint system for an ORC layout also contains OR-constraints. OR-constraints enable the designer to specify multiple alternative constraints, only one of which needs to be met, {\em e.g.}, a toolbar that is on the left side in landscape but is repositioned to the top (or bottom) of the screen in portrait mode.
The designer only needs to specify both (or any number of) alternative constraints joined with an OR-clause. 
Thus, with OR-constraints, a single layout specification can support many different screen sizes and aspect ratios.

We define OR-constraints through a mixture of hard and soft constraints. A whole ORC clause has a hard constraint, while each part of the OR-constraint has an independent soft constraint. This ensures layout specifications are always feasible. 
Some individual soft-constrained parts of an OR-constraint can be given higher priority than others through weights, which defines the level of importance or priority of that part.
For example, in the OR-constraint ``to the right OR at the beginning of the next row'', we likely prefer to place the current widget to the right of the previous one. Thus, we attach a larger, {\em e.g.}, double, weight to ``to the right'' part.


\subsection{ORC Editor}

The ORC Editor looks and works similar to other constraint-based GUI builders, with a canvas, palette, and properties panel \cite{zeidler2013auckland}. It supports safe use of ORC layout patterns (see Section \ref{sec:patterns}) through parameterised templates. Layouts can be edited interactively in the canvas through direct manipulation, or using the properties panel, {\em e.g.}, for template instantiation and modification. Designers can declaratively choose templates for widgets and widget containers and specify their parameters; the editor automatically creates and maintains the corresponding constraints. The template-based approach simplifies GUI maintenance compared to programming approaches \cite{myers2000past}, and a single flexible specification instead of several alternatives can ease authoring by reducing redundancy. As in ALE \cite{zeidler2013auckland}, users can still add and edit individual constraints, but we expect this to be used infrequently. Similar to ALE, the editor automatically solves the constraint system after each change and keeps the layout solvable by detecting, highlighting, and disabling conflicting constraints.

\subsection{Solving a System with OR-Constraints}

The naïve approach to solve a system with OR-constraints is to try all solutions. For very small systems, this might be feasible but does not scale well: for every OR-constraints with two parts,
there are two potential solutions. This number grows exponentially, {\em e.g.}, with only ten OR-constraints there are 1024 potential solutions that need to be evaluated. The naïve strategy quickly becomes impractical.

OR-constrained systems are a subset of SMT (Satisfiability Modulo Theory) problems. Instead of an explicit solver algorithm, we thus use Z3 \cite{de2008z3}, a very efficient SMT solver. Z3 typically scored highly in the yearly SMT competitions\footnote{SMT competitions: \url{smtcomp.org}} during the last decade.
One of the key insights of our work is that SMT solvers can be used to solve GUI layouts with OR-constraints.
Z3 has the ability to solve specifications with disjunctive sets of constraint alternatives, which makes it an ideal tool to solve a system of constraints that includes OR-constraints. Z3 chooses constraints by considering all possible combinations of alternatives and evaluating their appropriateness for the given situation with an objective function. The solver tries to satisfy as many constraints as possible, based on their assigned priorities. 
We have not experienced Z3 convergence failures. 
How constraints interact with each other can be controlled in two ways: 1) by setting constraint priorities, {\em e.g.}, prefer A over alternative A2, or 2) by combining constraints in different ways using Boolean operators, {\em e.g.}, (A AND B) OR (A2 AND B2) will force the solver to pick alternative branches in the same way.

Re-solving the entire constraint system iteratively after each change of a layout can be time-consuming and may, {\em e.g.}, slow down the speed of interactive GUI editing. To produce layout results dynamically at runtime, computations need to be efficient. By reusing parts of the previous solution, Z3 supports incremental solving, {\em i.e.}, it produces a result faster if the current constraint system is similar to the previous one. 
Instead of re-creating the whole constraint system after each adjustment in the layout, the ORC Editor removes obsolete constraints from the previously solved system and adds new ones.
This re-solves only part of the new constraint system based on the previous results and achieves better performance. 

\subsection{ORC Layouts for Screen Rotation}

\begin{figure}[t]
\includegraphics[width=\columnwidth]{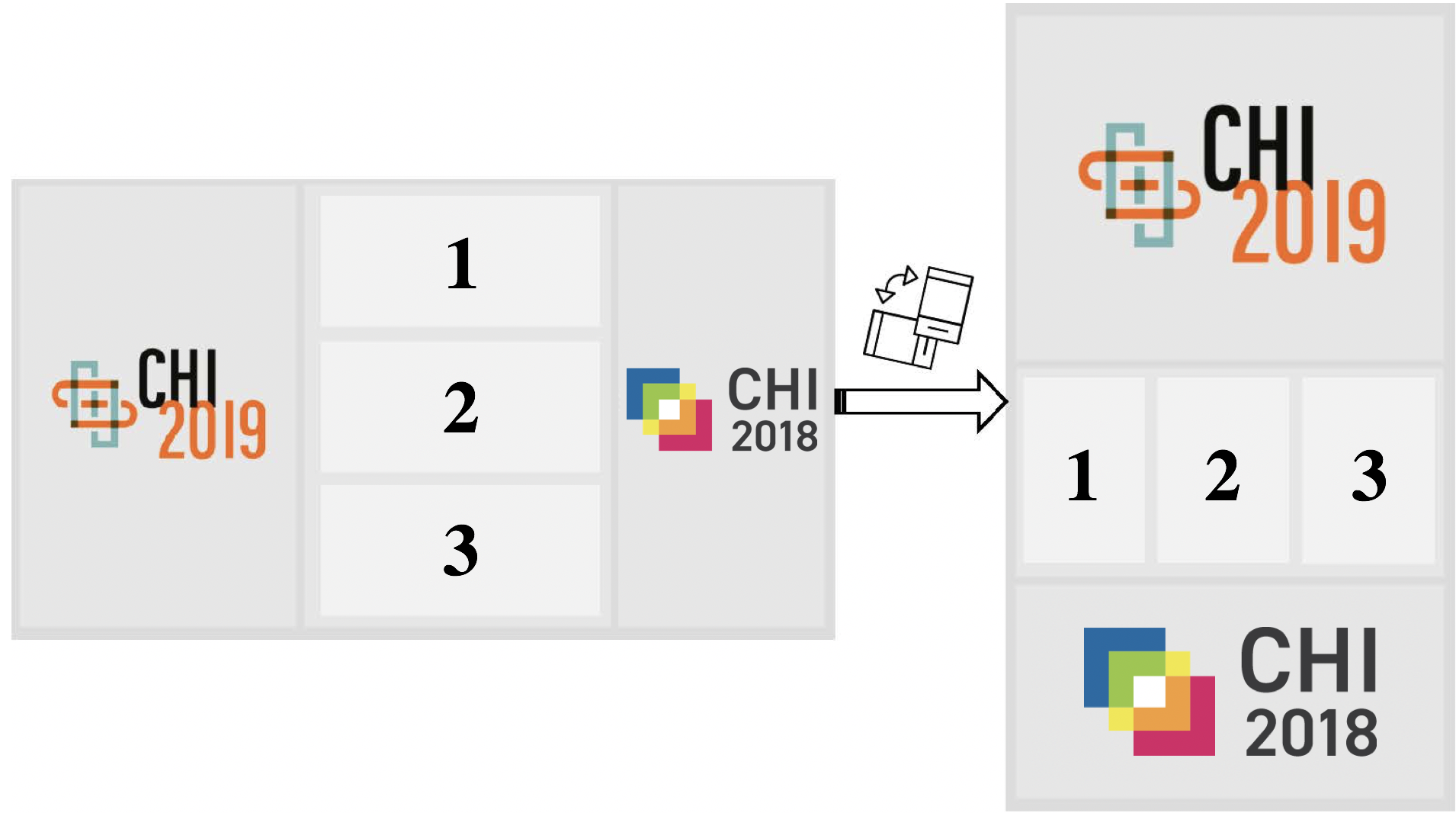}
\caption{An ORC layout with a single layout specification for horizontal and vertical rotation. Classic layouts, such as flow and grid, can either not specify the change in alignment between the widgets depending on the window aspect ratio, or cannot ensure consistent sizes across the different widget groups.}
\Description{This image shows an ORC layout with a single layout specification for horizontal and vertical rotation.}
\label{fig:rotation_example1}
\end{figure}

\begin{figure}[t]
\includegraphics[width=\columnwidth]{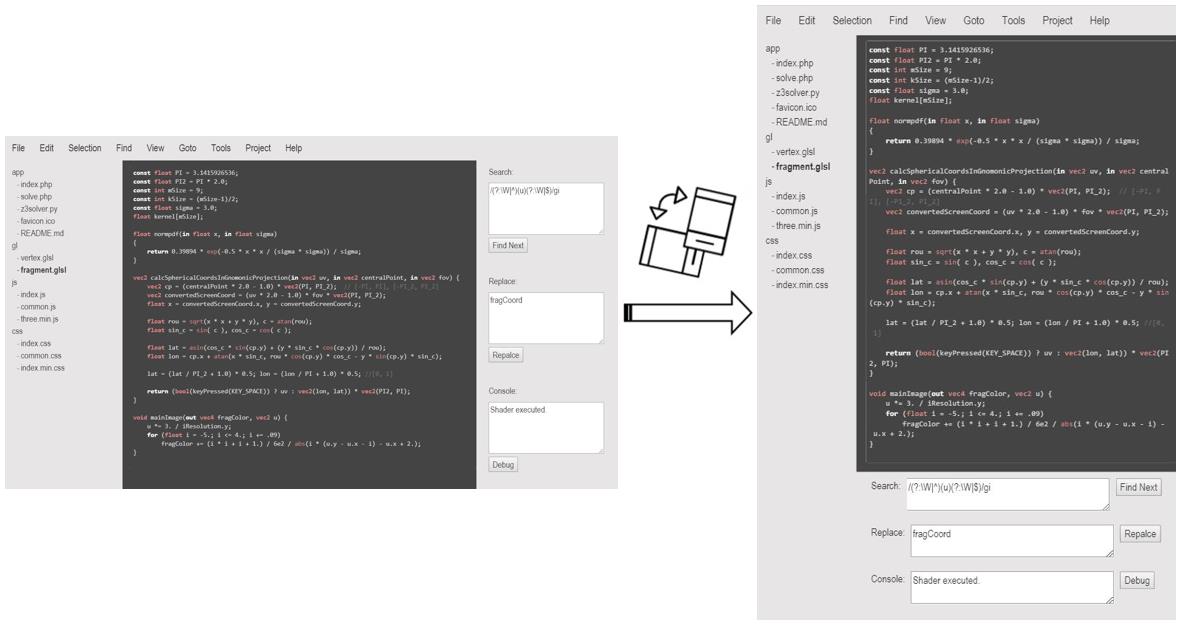}
\caption{A practical use case of ORC layouts for dealing with different device orientations. Note that the whole layout changes from horizontal to vertical while the inner layout of the control panel changes the opposite way. }
\Description{This image shows a practical use case of ORC layouts for dealing with different device orientations.}
\label{fig:rotation_example2}
\end{figure}

OR-constraints can unify flow layouts and grid layouts, which opens up options for layouts to work across screen rotations and cases where grid and flow layouts individually fail. 
This enables designers to seamlessly deal with screen rotations with a single layout specification.
For instance, neither grid nor flow layouts can deal with the case in \autoref{fig:rotation_example1} through a single layout specification.
Although the general change in topology may be achieved using two nested flow layouts, the flow layouts cannot ensure that the widgets are consistently aligned as shown.
There are three high-level widgets (labeled areas) in the layout: CHI 2019, block, and CHI 2018, denoted as [a], [b], and [c] respectively. Inside block [b], we have three low-level widgets: [1], [2], and [3]. Using ORC, the alignments in the two alternatives (horizontal and vertical) can be specified separately using constraints and then combined with OR. In contrast to other layout methods, ORC can also ensure consistent sizes across the two groups [a], [b], [c] and  [1], [2], [3] as follows:

\begin{align*}
&\left(
    \left(
        \MB_\HEIGHT = \sum_{i=1}^3 \MI_\HEIGHT 
    \right)
    \wedge
    \left(
        \MB_\WIDTH = \max_{i=1,2,3} \MI_\WIDTH 
    \right)
\right)
\\
{\bf OR}
&\left( 
    \left(
        \MB_\WIDTH = \sum_{i=1}^3 \MI_\WIDTH 
    \right)
    \wedge
    \left(
        \MB_\HEIGHT = \max_{i=1,2,3} \MI_\HEIGHT 
    \right)
\right)
\end{align*}
The two soft constraints above specify the correct alignment for the horizontal case (\autoref{fig:rotation_example1} left) or the vertical case (right), respectively.
A more practical use case of an ORC layout for device rotation is shown in \autoref{fig:rotation_example2}. This layout contains a menu, a file explorer panel, a main view for code editing, and a control panel with widgets for searching, replacing, and console debugging. It uses OR-constraints similar to the previous example. 

\section{ORC Layout Patterns} \label{sec:patterns}

ORC layout covers a superset of the most popular layout mechanisms, allowing designers to ``blend'' them. To achieve this in a controlled manner, we propose patterns as basic building blocks that can be combined into layouts of arbitrary complexity. For example, ORC layout patterns can specify constraints for cross-cutting (Figures \ref{fig:teaser}), connecting (Figure \ref{fig:connected_cross_cutting}), further constraining (Figures \ref{fig:balanced} and \ref{fig:around}) and re-positioning sub-layouts (Figures \ref{fig:rotation_example1}-\ref{fig:h_v} and \ref{fig:alternative}), or replacing them entirely (Figure \ref{fig:optional}). Furthermore, multiple patterns can be combined (Figure \ref{fig:new_ex}).
Here we present the constraints for several of our new layout patterns. To our knowledge no current layout mechanism or editor can deal with the shown examples, except through multiple layout specifications or code.


\subsection{Mixed Layout Specifications}

Previous layout managers often focus on solving a single type of layout. The use of OR-constraints makes it possible to mix different types of constraints and layout specifications. For example, a ``to the right OR at the start of the next row below'' constraint between each successive pair of widgets enables ORC layouts to unify flow and grid-bag layouts. With such a unified layout, the designer only needs to create a single layout specification for both landscape and portrait modes. 


In the following, we assume that all widgets are ordered in the layout. We denote the $i$-th widget as $\MI$ and the set of all widgets as $W$. 
For each widget $\MI \in W$, we define its left \& top position, width, and height as $\MI_\LEFT$, $\MI_\TOP$, $\MI_\WIDTH$, $\MI_\HEIGHT$ respectively. 
The soft constraint ``to the right'', $C_{Right}$, is  formulated as:

\begin{align*}
    C_{Right} & := \left( \MI_\LEFT = \MII_\LEFT+ \MII_\WIDTH \right) \\
    & \wedge \left( \MI_\TOP = \MII_\TOP \right)
\end{align*}

The soft constraint ``at the start of the next row below'', $C_{NextRow}$, is formulated as:

\begin{align*}
C_{NextRow} &:= (\MI_\LEFT = 0)  \wedge \\
 & \left( \MI_\TOP\geq \MJ_\TOP + \MJ_\HEIGHT, \forall j < i, \MJ \in W  \right)
\wedge \\
& \left( \bigvee_j \MI_\TOP = \MJ_\TOP + \MJ_\HEIGHT, \forall j < i, \MJ \in W \right)
\end{align*}

\begin{figure}[t]
\includegraphics[width=\columnwidth]{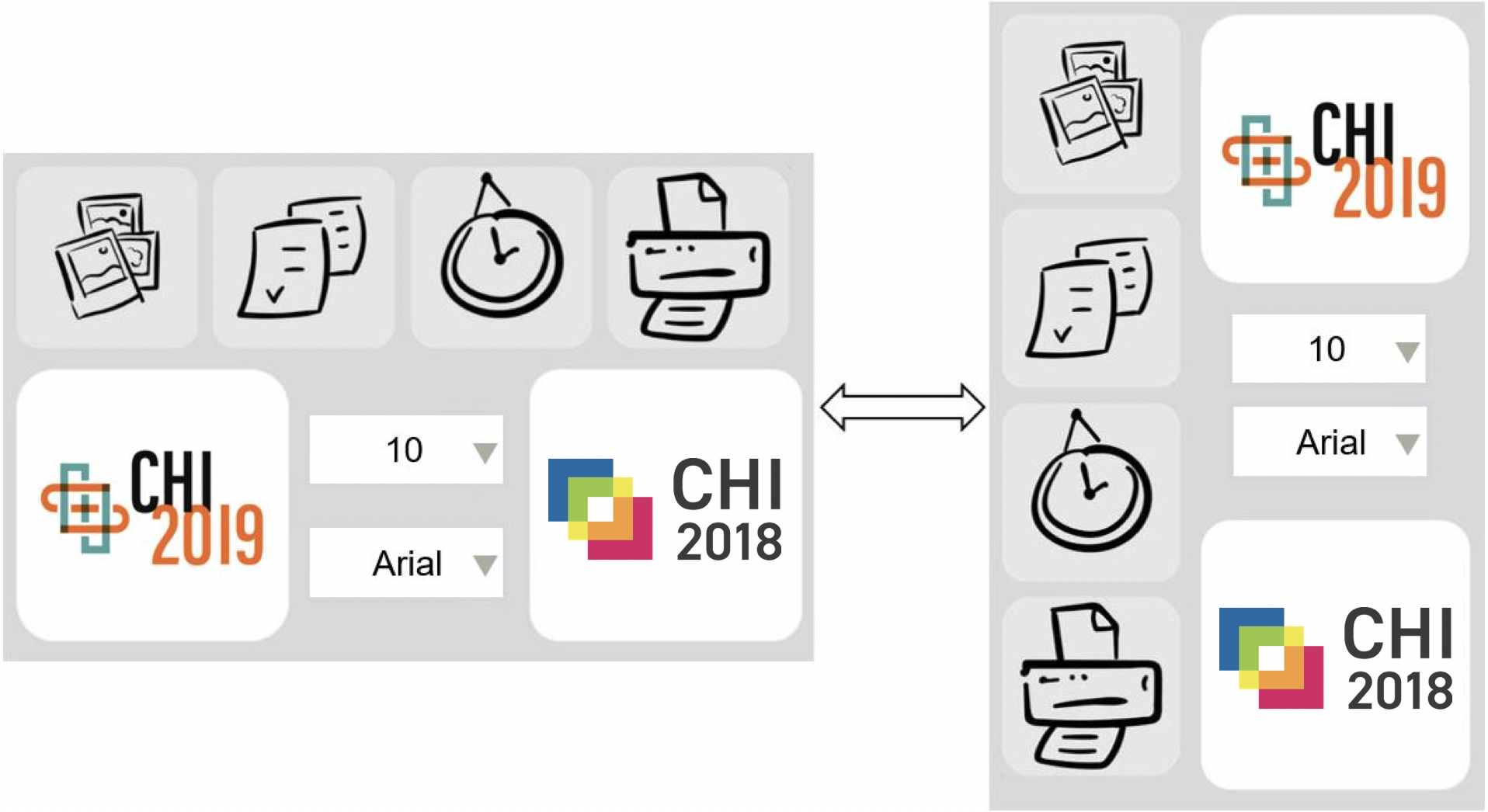}
\caption{The horizontal window uses a ``to the right OR at the start of the next row'', while the vertical one has an OR-constraint ``to the bottom OR at the start of the next column''.}
\Description{This image shows an example of switching between the horizontal flow layout pattern and the vertical flow layout pattern.}
\label{fig:h_v}
\end{figure}

The OR-constraint $C_{Horizontal}$ expressing horizontal flow layout is then a hard constraint (or one with very high weight) with a disjunction of the two soft constraints ``to the right'' and ``at the start of the next row below'':

\begin{equation*}
    C_{Horizontal} := C_{Right}\ \ OR\ \ C_{NextRow}
\end{equation*}
Vertical flow layouts use corresponding OR-constraints for ``to the bottom OR at the start of the next column''. With ORC, the choice of using a horizontal or vertical flow layout can even be left to the constraint solver for flexibility by combining both alternatives with an OR, as illustrated in \autoref{fig:h_v}.


\subsection{Cross-Cutting Layout Pattern}


The cross-cutting layout pattern achieves a consistent appearance in a GUI by adding constraints that cross-cut the boundaries between the different sub-layouts of a GUI. This is illustrated in \autoref{fig:teaser} (b), where each widget in the top toolbar has the constraint ``to the right OR at the beginning of next row'' and each in the left toolbar has the constraint ``to the bottom OR at the beginning of next column''. The toolbars automatically transform into multiple rows/columns as the screen gets smaller, while cross-cutting constraints ensure that the widgets in the toolbars have sizes consistent with one another.

\subsection{Connected Layout Pattern}

\begin{figure}[t]
\includegraphics[width=\columnwidth]{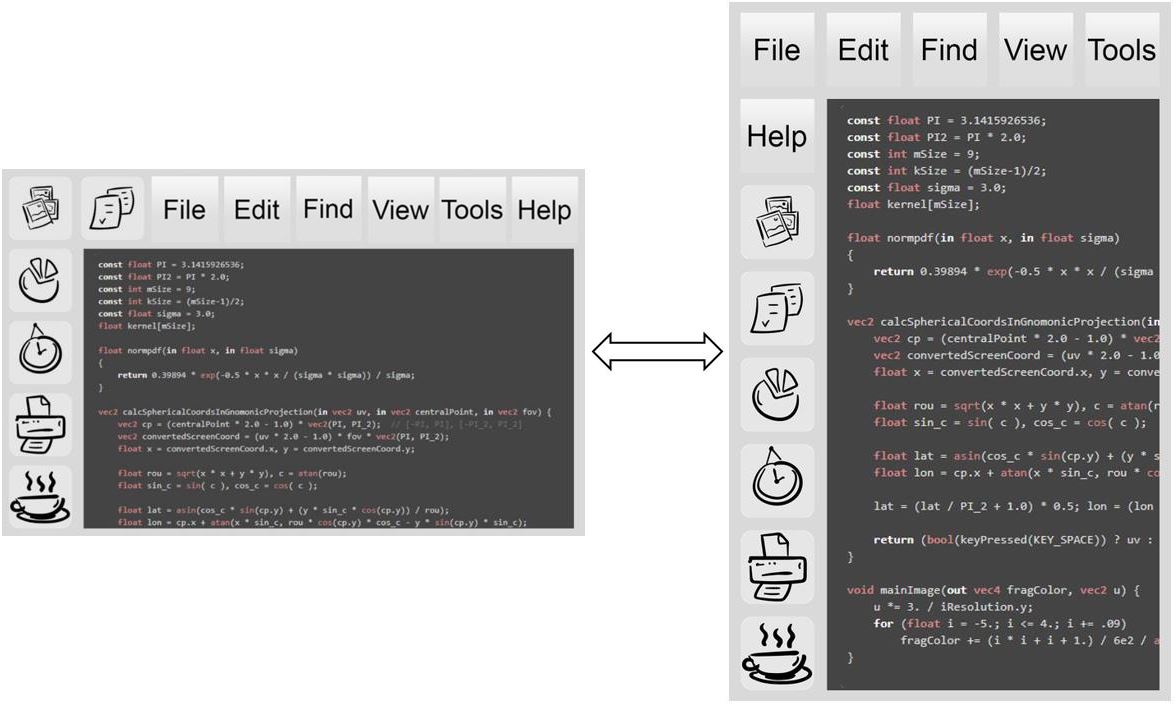}
\caption{Connected layout pattern:  the left and top toolbars are connected so widgets can move between them.}
\Description{This images shows connected layout pattern. Two toolbars are at the top and to left of the text area, respectively.}
\label{fig:connected_cross_cutting}
\end{figure}

The connected layout pattern is a generalization of different layouts, such as horizontal and vertical flow, which emerges by connecting the sub-layouts of a GUI so that widgets can move between them. This can be achieved by refining the constraints that move widgets into a new position based on available space, as discussed for flow layouts in the section about mixed layout specifications. For example, the horizontal and vertical toolbar areas in \autoref{fig:connected_cross_cutting} are connected to create a better fit in the available screen space. The most suitable number of widgets in the top toolbar is $t_{best} := window\_width\ /\ widget\_width$. If the original number of widgets in the top toolbar $t$ is smaller than $t_{best}$, then the first $t_{best}-t$ widgets in the left toolbar are moved to the top one. Otherwise, the first $t-t_{best}$ widgets in the top toolbar are moved to the left.

\subsection{Balanced Flow Pattern}

\begin{figure}[t]
\includegraphics[width=\columnwidth]{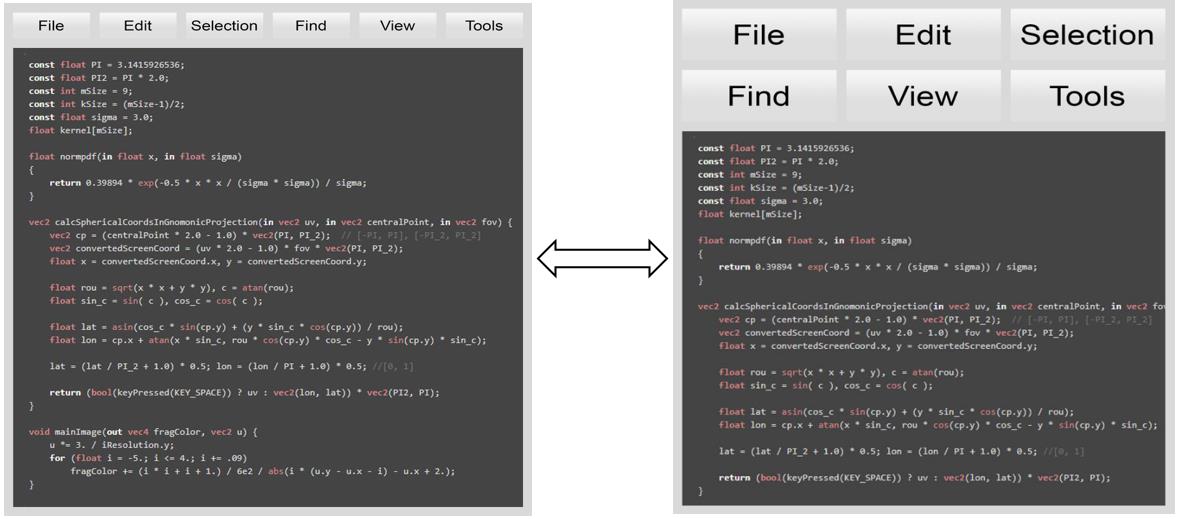}
\caption{Balanced flow pattern: as the toolbar widgets are broken into rows, each row can only have 1, 2, 3, or 6 widgets.}
\Description{This images shows balanced flow pattern. A toolbar with 6 widgets are at the top of the text area.}
\label{fig:balanced}
\end{figure}

Conventional flow layouts do not ensure that their rows or columns are balanced with respect to the contained widgets. ORC layouts support balanced layout alternatives and combine them with an OR. For example, when the widgets in the toolbar in \autoref{fig:balanced} are broken into rows, each row can only have a predefined number of widgets, which is a factor of the number of the widgets in the toolbar. We compute the set of all the factors of the number of the widgets in the toolbar. According to the window size, the top toolbar has a preferred size and number of rows. Choosing the closest number $c$ in the factor set, $t/c$ widgets are placed in each row. 

\subsection{Alternative Positions Pattern}

\begin{figure}[t]
\includegraphics[width=\columnwidth]{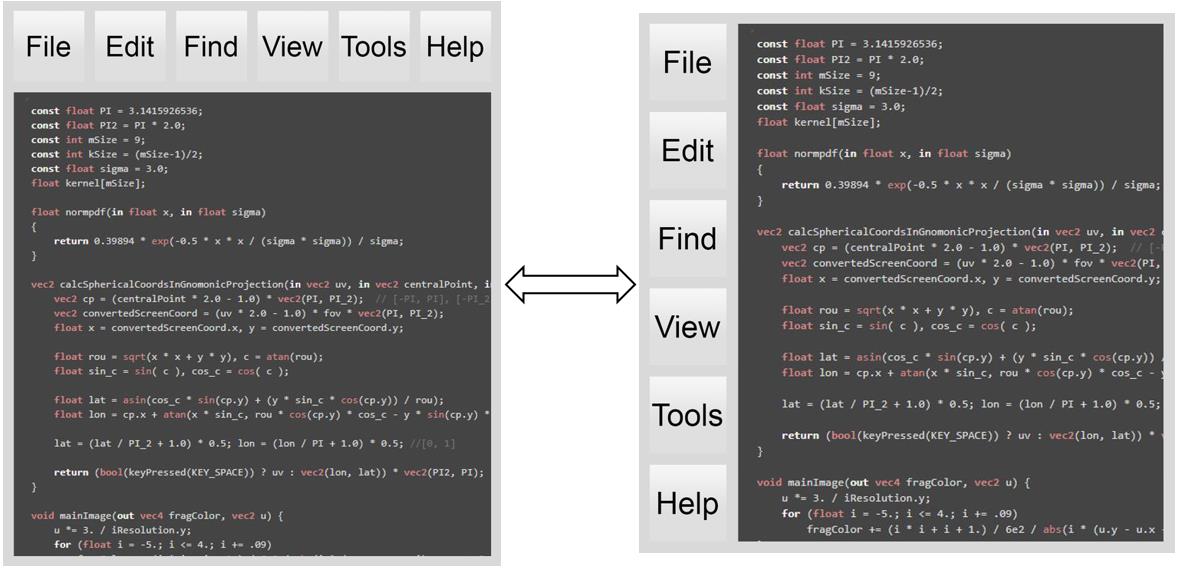}
\caption{Alternatives positions pattern: the toolbar can be placed on top OR to the left.}
\Description{This image shows alternative positions pattern. There is a toolbar besides the text area, which can be placed on top OR to the left.}
\label{fig:alternative}
\end{figure}

With a simple OR-constraint, we can define alternative positions for widgets or entire sub-layouts. For example, the top toolbar in \autoref{fig:alternative} can be changed to be a left toolbar when the screen size changes or vice versa.

\subsection{Alternative Widgets Pattern}

If a window does not have enough space, some widgets can have alternative representations. For instance, \textit{List Boxes} can be replaced by \textit{Option Menus}. This is done by putting both of them into the layout, next to each other, but making one of them invisible by giving it a size of 0. We express this through constraints as follows, where the weight of (1) > (2):

\begin{quote}
$(size(ListBox) = prefSize(ListBox)$\\ \quad $\land  size(OptionMenu) = 0 ) \quad (1)$\\
OR $(size(ListBox) = 0$\\
\quad $\land size(OptionMenu) = prefSize(OptionMenu)) ~~ (2)$
\end{quote}

\subsection{Optional Layout Pattern}
\begin{figure}[t]
\includegraphics[width=\columnwidth]{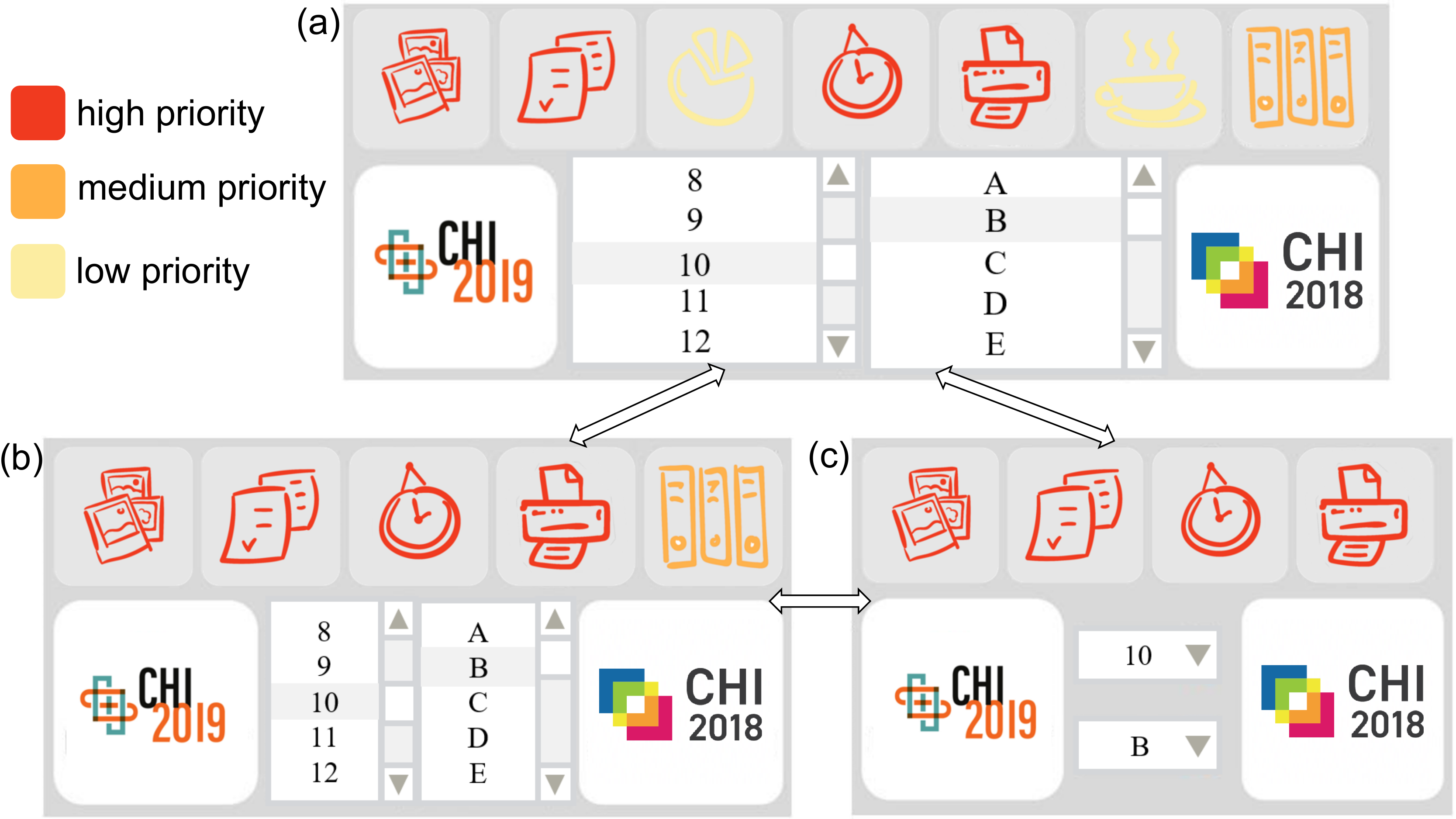}
\caption{Optional layout and alternative widgets pattern example. (a) shows a full-sized window with all the widgets visible. (b) shows a medium-sized window where widgets with low weights disappear. (c) further reduced window size so that widgets with medium and low weights disappear. Both radio buttons and the list box are replaced by option menus to save space.}
\Description{This image shows an optional layout and alternative widgets pattern example.}
\label{fig:optional}
\end{figure}

\begin{figure*}[t]
\includegraphics[width=0.7\textwidth]{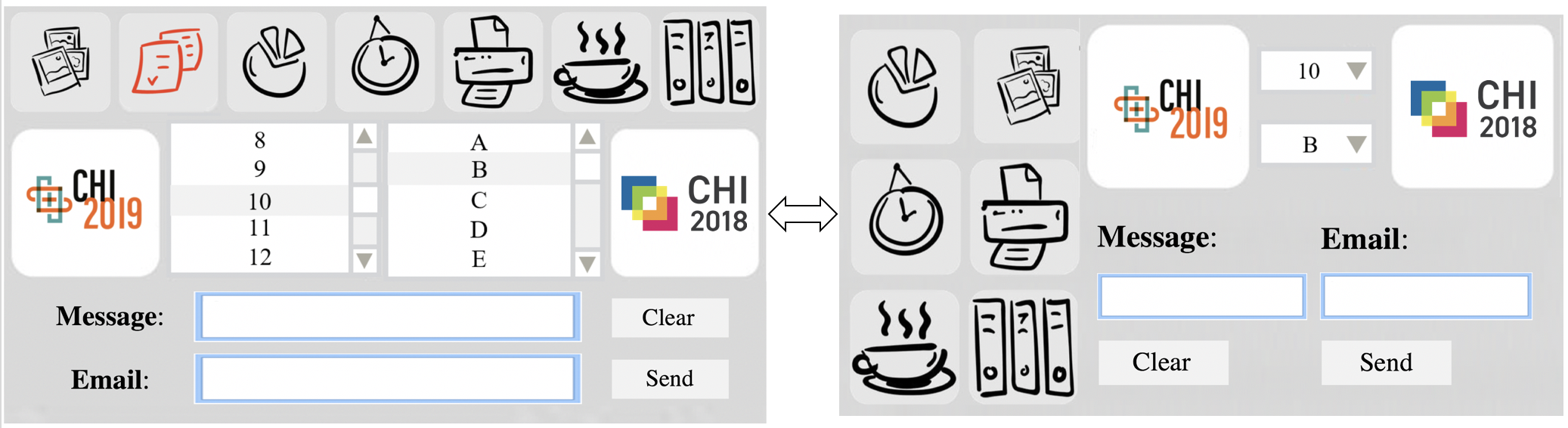}
\caption{A more complex example combining the alternative position pattern for the toolbar at the top and the widgets at the bottom, the balanced flow and optional layout patterns for the toolbar, and the alternative widgets pattern for the list boxes.}
\Description{This image shows a more complex example combining multiple layout patterns.}
\label{fig:new_ex}
\end{figure*}

When a window is made smaller, optional widgets might disappear, like in the ribbon menu in MS Word. Similar to the alternative widgets pattern, the OR-constraint ``the widget has zero size OR nonzero size'' realizes this functionality. 
For example, each widget in the ribbon menu is assigned a priority through its weight (\autoref{fig:optional}, \autoref{fig:new_ex}).
Thus, when the window gets smaller, widgets with lower weights disappear before those with higher weights. We use the following constraints to implement this:
    $size = prefSize\ OR\ size = 0.$

For widgets with high priority, $size = prefSize$ is a hard constraint (or one with a very large weight). For widgets with medium priority, $size = prefSize$ is a soft constraint with a large weight $A$, and $size = 0$ is a soft constraint with a small weight $B$. For widgets with low priority, $size = prefSize$ is a soft constraint with a large weight $C$, and $size = 0$ is a soft constraint with a small weight $D$,
where weight $A$ > weight $C$ and  weight $B$ < weight $D$.

The example in \autoref{fig:optional} shows both the widget alternative and optional functionalities. It demonstrates how the ribbon menu changes as the window gets smaller. Widgets in the ribbon menu disappear according to their weights as the window gets smaller. In the bottom row, list boxes and radio buttons are replaced by option menus to save space.


\subsection{Flowing Widgets around a Fixed Area}

\begin{figure}[t]
  \includegraphics[width=\columnwidth]{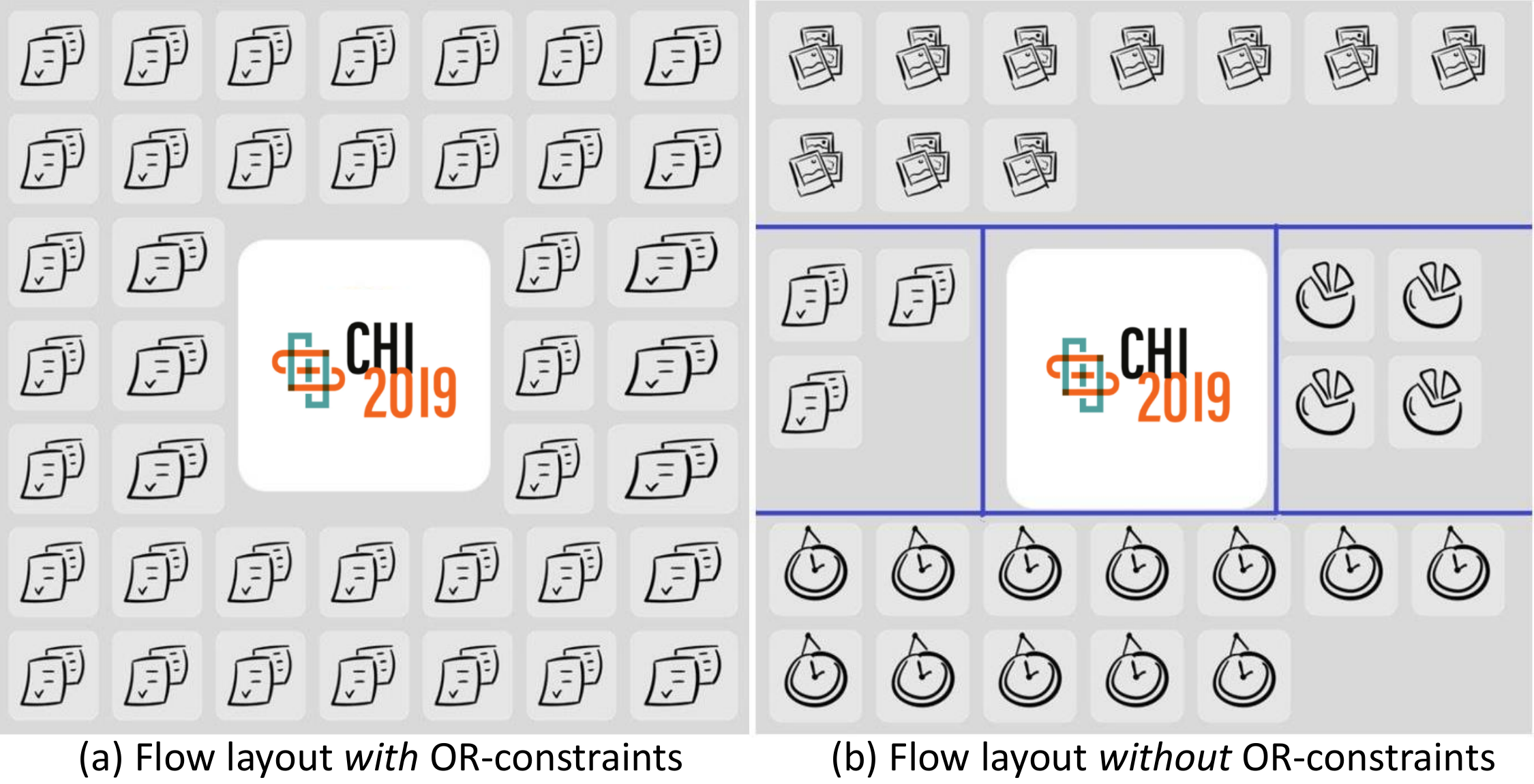}
  \caption{(a) Flow layout with OR-constraints.  The position of the ``CHI 2019'' image is fixed while all the other widgets flow around it. The size of the flowing widgets adjusts to fill up the available space. 
  (b) Flow layout without OR-constraints.}
  \Description{This image shows a flow layout with an image fixed in the middle while all the other widgets flow around it.}
  \label{fig:around}
\end{figure}

An OR-constraint system can even deal with the case of flowing widgets around a fixed area, {\em e.g.}, similar to placing a picture on a page of text and flowing text around it. As shown in \autoref{fig:around} (b), with simple flow layouts a fixed widget would necessitate splitting the window into 4 parts and widgets would flow in each part separately, which results in an undesirable outcome.
However, with OR-constraints, we can flow widgets around a fixed area without splitting the window into parts. The OR-constraint system can solve the whole window at the same time to arrive at a much more pleasing result and even resize widgets to avoid ``ragged right'' margins.

If a widget is above or below the fixed area, then it has the OR-constraint ``to the right of the previous widget OR at the beginning of next row'' as usual. When a row of widgets overlaps with the rows occupied by the fixed area, this will cause a row break to prevent overlap. In essence, the fixed widget cuts the rows into two parts, as in the connected layout pattern. Algorithm \ref{alg:around} shows how we implemented this layout. The algorithm consists entirely of conditionals and constraints with varying weights, and can therefore be translated into Boolean logic as supported by ORC.
$W \rightarrow L$ is shorthand for a position constraint that places a widget $W$ into location $L$. ``if A then B'' equals the constraint ``NOT(A) OR B''.

\begin{algorithm}[t]

    \SetKwInOut{Input}{Input}
    \SetKwInOut{Output}{Output}
    
\SetAlgoLined

first widget $\rightarrow$ top left corner

\If {overlap} {
first widget $\rightarrow$ after fixed area}

\If {current widget is above or below fixed area} {
    	current\_left = prev\_right {\bf OR} current $\rightarrow$ next\_row 
    	\textcolor{blue}{\ (hard)}
        }
\If {current widget shares any row with the fixed area}
{
    \tcp {break over rows to prevent overlap}
    \If {there is enough space for current widget in this row} 
    {
        current\_top = prev\_top \textcolor{blue}{\ (hard)}\\
        \If {enough space between previous widget and fixed widget} {
            current\_left = prev\_right
            \textcolor{blue}{\ (hard)}
        } \Else {
             current\_left = fixed\_right
            \textcolor{blue}{\ (hard)}
        }
    }
    \Else{
	    current $\rightarrow$ next\_row\\	 
    	\If {enough space between previous widget and fixed area} {
            current\_left = left\_boundary
            \textcolor{blue}{\ (hard)}
        }
    	\Else {
            current\_left = fixed\_right
            \textcolor{blue}{\ (hard)}
        }
    }
}

current\_top = prev\_top \textcolor{blue}{\ (soft with high weight)}

current $\rightarrow$ next\_row\textcolor{blue}{\ (soft with low weight)}

\caption{Flowing widgets around a fixed area.}
 
\label{alg:around}
\end{algorithm}

Document layout methods also address this case, but with completely different methods. 
For example, Adaptive Grid-Based Document Layout \cite{Jacobs2003document} uses adaptive layout templates to implement similar functionality, by defining sets of constraints for different cases.
With ORC layout, all cases can be solved through a single constraint system.

\section{Implementation}

We implemented our prototypical ORC Editor in Python 2.7 with Tkinter, a GUI library for Python, and the Z3-solver, where we used the default WMax optimization algorithm for weighted soft constraint problems \cite{nieuwenhuis2006wmax}.

\section{Evaluation}

Efficiency plays a vital role in solving constraint-based systems. We measured the execution time of the ORC layout solver for different layout editing operations (insert, delete, move, and resize) with different numbers of constraints. For each condition, we measure the average execution time of 10 runs. We conduct the experiments on a laptop with an Intel i5 CPU. As shown in Table 1, with fewer than 175 constraints our method can efficiently solve the most common cases in less than one second on this average machine.
For deployment, one could cache solved layouts for typical devices to improve performance. Note that for insert, delete, and move, our ORC Editor incrementally updates the constraint system, which saves approximately 40\%-86\% of the computation time for 5 widgets, and around 30\%-48\% of the time for 20 widgets. In contrast, a resize operation typically requires a complete rebuild of the entire constraint system, which is costlier.

\begin{table}[t]
    \centering
    \includegraphics[width=\columnwidth]{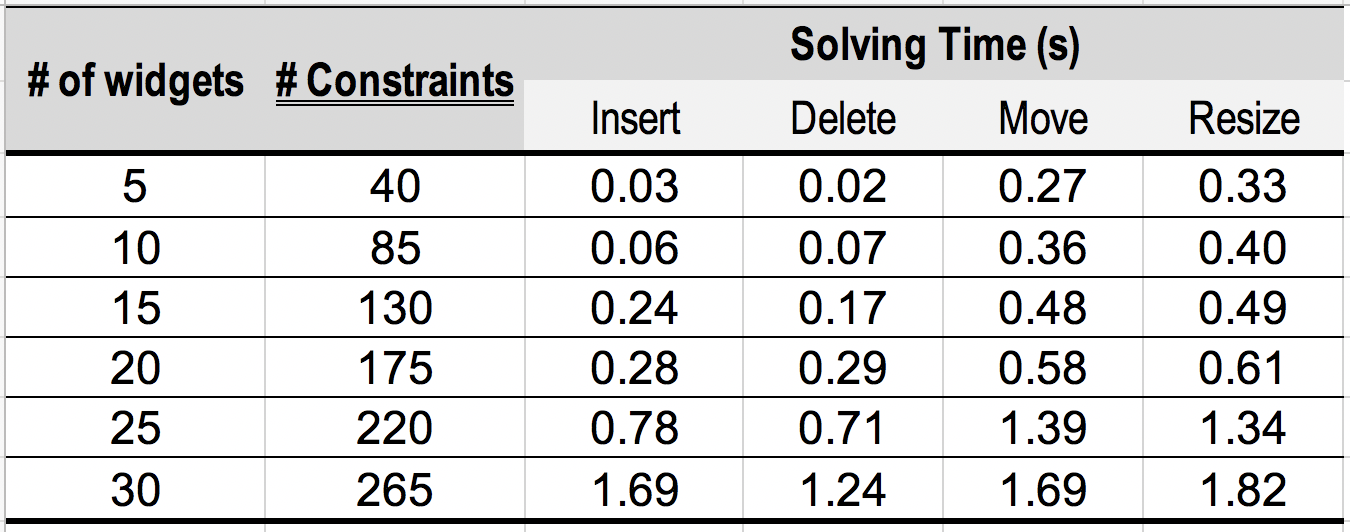}
    \caption{Average solving times with ORC layout (in seconds) for different number of constraints and different operations. (“Resize” in this table means resizing specific widgets.) In our experiments, we used horizontal flow layout with the OR-constraint ``to the right OR at the beginning of next line'' and the sizes of widgets were determined by their min/pref/max constraints.}
    \label{tbl:timing}
\end{table}

Although ORC solving does not yet work at real-time interactive speeds, layouts can be solved during app loading, pre-computed in a background thread, and cached for typical screen sizes. For more complex cases, our method could also take advantage of hierarchical layouts to efficiently solve a new layout in under one second. 
For example, smaller widgets could be contained in larger bounding boxes, and the layout of individual boxes could be solved efficiently in parallel. 

\subsection{Expert Review}

We initially considered a comparison to see how ORC layouts perform relative to other methods. Yet, to our knowledge, there are currently no GUI layout systems that can create a single easily ``rotatable'' layout that works for non-trivial examples. Thus, we started by exploring if designers could work with ORC layouts by performing a qualitative evaluation with nine participants (four female, aged from 21 to 48). All  participants were professional UI designers or software engineers from technical companies with substantial experience with GUI layouts and responsive web design.

\subsubsection{Procedure}

We first showed participants PowerPoint slides with all the figures in this paper and asked them to follow the sequence of edit operations for those figures to  familiarize them with our ORC Editor. All the participants found this easy to follow. One stated that ``
{\em I think anyone who has ever used a responsive web app will find it fairly intuitive.}''
It took only about five minutes for them to finish trying all the edit operations. After having experienced the ORC Editor, most participants were curious about the difference between ``responsive design'' and ORC layouts. When told that our layout method enables the designer to create only a single layout specification for multiple screen sizes and orientations, a UI designer stated that {\em``Seems like a great idea. [ORC layouts] could save us a lot of time.''} 

Then we asked them to create several layouts containing 8-12 widgets in a single container that work well in landscape and portrait mode with the ORC Editor, and interviewed them about their opinions, including how well layouts worked for different screen orientations. In general, they quickly grasped how ORC layout patterns work and were able to create new, rotatable layouts in about 10-20 minutes. 

\subsubsection{Interviews}

In post-hoc interviews, participants expressed generally very positive feelings towards ORC layouts. They found ORC layout editing user-friendly, and many compared their experience with ORC favorably with their day-to-day experience of building responsive UIs. They identified that this new layout method complements existing ones and indicated that {\em``[ORC layouts] allows me to design layouts that are otherwise hard to do. The results of the layout did meet my expectation[s]''}. One stated that {\em``Usually we need to carefully design different layouts for phones, tablets, and PCs. But [ORC] has great potential to automate [such] repetitive work. [ORC layouts] makes it so much simpler to deal with [mobile device] layouts, as I then need to maintain only one layout''}. Some usability issues of the ORC Editor prototype were criticized, but participants also recognized that integration into an industry-strength GUI editor would address such issues.

\section{Discussion}

As demonstrated above, OR-constraints enable new ways to specify GUI layouts that can automatically adapt to different screen orientations and sizes. Relative to previously presented layout methods, OR-constraints can express more extensive re-arrangements of widgets. In particular, OR-constraints are able to combine the power of both flow and grid-based layout specifications. A big benefit of ORC layouts is that this method enables the designer to create only a single layout specification that works for multiple screen sizes and orientations. This means that designers do not need to create and maintain multiple layout specifications for each GUI screen.

We explain the trade-off between power and simplicity in the ORC editor as follows: When a user specifies a layout for a GUI container, the system creates the corresponding constraints with default parameters. Through our pre-defined constraint layout patterns, we substantially reduce the complexity presented to the user. Users then only need to modify the parameters and select different options for a layout functionality (but not the constraints themselves) to meet more specialized requirements as desired. By automatically keeping layouts solvable \cite{zeidler2013auckland}, we keep layout editor usability high, as this effectively prevents users from over-specifying a layout. 

It is advantageous to create a single layout specification for multiple different ({\em e.g.}, portrait and landscape) form factors. Similar to challenges in software engineering, it is easy for multiple alternative GUI layout specs to get out of ``sync'', {\em e.g.}, when widgets get added/changed/removed. While there may be other ways to deal with this situation, it typically requires decomposing a GUI into many individual building blocks. A single spec is easier to maintain and to keep consistent. 

\subsection{Limitations}
One issue with ORC layouts is that for more complex situations a hierarchy may be necessary to solve the constraint system. 
While a single constraint system offers potentially more flexibility, some naïve forms of prioritization of high-level vs.\ low-level widgets can make a constraint system time-consuming to solve. As GUI designers are used to decompose complex layouts into a hierarchy we currently do not see this as a major issue.
We acknowledge that usability, layout, and aesthetic design principles are not directly supported by the work reported here. If such aspects are expressible as constraints, they can be specified as part of an ORC layout. Yet, we consider this outside the scope of our current work.

\section{Conclusion}
In this paper, we presented ORC layouts, a novel GUI layout method that adds OR-constraints to standard constraint-based layout specifications. 
ORC layout unifies grid layout and flow layout, offering new possibilities for flexible GUIs that are not supported by any other layout method.
Through the use of an efficient SMT solver, ORC layout constraint systems with soft, hard and OR-constraints can typically be solved at interactive rates. 
We envision that our new layout method could be widely applied in modern web design, document format, and app layout. We plan to open-source our code for future research on OR-constraint layouts.

Z3 is a very general constraint solver. Likely, this generality introduces some computation overhead. As our work is only the first exploration of ORC layouts, we envision that future layout solvers will improve performance further.
Thus, we see potential for new, more specialized solving algorithms that can deal with OR-constraints in a more efficient manner.

\subsection{Acknowledgements}
This work builds on initial explorations of OR-constrained layouts by Navid Mohaghegh and the last author.

\bibliographystyle{ACM-Reference-Format}
\balance
\bibliography{ORCLayouts.bib}

\end{document}